\documentclass[aps,prl,reprint,longbibliography,preprintnumbers,floatfix,superscriptaddress]{revtex4-2}
\usepackage{graphicx} 
\usepackage{dcolumn}
\usepackage{bm}
\usepackage{amsmath}
\usepackage{multirow}
\usepackage{placeins} 
\usepackage{soul}
\usepackage{booktabs} 
\usepackage{xspace, tabularx}
\usepackage[caption=false]{subfig}
\usepackage{accents}
\usepackage{xcolor}
\usepackage{placeins}
\usepackage{threeparttable}
\usepackage{lineno}
\usepackage{rotating}
\usepackage{units}

\makeatletter
\newcommand{\oset}[3][0ex]{%
  \mathrel{\mathop{#3}\limits^{
    \vbox to#1{\kern-2.5\ex@
    \hbox{$\scriptstyle#2$}\vss}}}}
\makeatother

\newcommand{\papertitle}{Determining Neutrino Mass Ordering with NOvA and Upcoming JUNO Measurements}

\usepackage{subfiles} 

\begin{document}

\title{\papertitle}
\preprint{FERMILAB-PUB-26-0387-PPD}

\newcommand{\ANL}{Argonne National Laboratory, Argonne, Illinois 60439, 
USA}
\newcommand{\Bandirma}{Bandirma Onyedi Eyl\"ul University, Faculty of 
Engineering and Natural Sciences, Engineering Sciences Department, 
10200, Bandirma, Balıkesir, Turkey}
\newcommand{\ICS}{Institute of Computer Science, The Czech 
Academy of Sciences, 
182 07 Prague, Czech Republic}
\newcommand{\IOP}{Institute of Physics, The Czech 
Academy of Sciences, 
182 21 Prague, Czech Republic}
\newcommand{\Atlantico}{Universidad del Atlantico,
Carrera 30 No.\ 8-49, Puerto Colombia, Atlantico, Colombia}
\newcommand{\BHU}{Department of Physics, Institute of Science, Banaras 
Hindu University, Varanasi, 221 005, India}
\newcommand{\UCLA}{Physics and Astronomy Department, UCLA, Box 951547, Los 
Angeles, California 90095-1547, USA}
\newcommand{\Caltech}{California Institute of 
Technology, Pasadena, California 91125, USA}
\newcommand{\Cochin}{Department of Physics, Cochin University
of Science and Technology, Kochi 682 022, India}
\newcommand{\Charles}
{Charles University, Faculty of Mathematics and Physics,
 Institute of Particle and Nuclear Physics, Prague, Czech Republic}
\newcommand{\Cincinnati}{Department of Physics, University of Cincinnati, 
Cincinnati, Ohio 45221, USA}
\newcommand{\CSU}{Department of Physics, Colorado 
State University, Fort Collins, CO 80523-1875, USA}
\newcommand{\CTU}{Czech Technical University in Prague,
Brehova 7, 115 19 Prague 1, Czech Republic}
\newcommand{\Dallas}{Physics Department, University of Texas at Dallas,
800 W. Campbell Rd. Richardson, Texas 75083-0688, USA}
\newcommand{\DallasU}{University of Dallas, 1845 E 
Northgate Drive, Irving, Texas 75062 USA}
\newcommand{\Delhi}{Department of Physics and Astrophysics, University of 
Delhi, Delhi 110007, India}
\newcommand{\JINR}{Joint Institute for Nuclear Research,  
Dubna, Moscow region 141980, Russia}
\newcommand{\Erciyes}{
Department of Physics, Erciyes University, Kayseri 38030, Turkey}
\newcommand{\FNAL}{Fermi National Accelerator Laboratory, Batavia, 
Illinois 60510, USA}
\newcommand{\FSU}{Florida State University, Tallahassee, Florida 32306, USA}
\newcommand{\UFG}{Instituto de F\'{i}sica, Universidade Federal de 
Goi\'{a}s, Goi\^{a}nia, Goi\'{a}s, 74690-900, Brazil}
\newcommand{\Guwahati}{Department of Physics, IIT Guwahati, Guwahati, 781 
039, India}
\newcommand{\Harvard}{Department of Physics, Harvard University, 
Cambridge, Massachusetts 02138, USA}
\newcommand{\Houston}{Department of Physics, 
University of Houston, Houston, Texas 77204, USA}
\newcommand{\IHyderabad}{Department of Physics, IIT Hyderabad, Hyderabad, 
502 205, India}
\newcommand{\Hyderabad}{School of Physics, University of Hyderabad, 
Hyderabad, 500 046, India}
\newcommand{\IIT}{Illinois Institute of Technology,
Chicago IL 60616, USA}
\newcommand{\Imperial}{Imperial College London, Department of Physics,
 London, United Kingdom}
\newcommand{\Indiana}{Indiana University, Bloomington, Indiana 47405, 
USA}
\newcommand{\INR}{Institute for Nuclear Research of Russia, Academy of 
Sciences 7a, 60th October Anniversary prospect, Moscow 117312, Russia}
\newcommand{\UIowa}{Department of Physics and Astronomy, University of Iowa, 
Iowa City, Iowa 52242, USA}
\newcommand{\ISU}{Department of Physics and Astronomy, Iowa State 
University, Ames, Iowa 50011, USA}
\newcommand{\Irvine}{Department of Physics and Astronomy, 
University of California at Irvine, Irvine, California 92697, USA}
\newcommand{\Jammu}{Department of Physics and Electronics, University of 
Jammu, Jammu Tawi, 180 006, Jammu and Kashmir, India}
\newcommand{\Lebedev}{Nuclear Physics and Astrophysics Division, Lebedev 
Physical 
Institute, Leninsky Prospect 53, 119991 Moscow, Russia}
\newcommand{\Magdalena}{Universidad del Magdalena, Carrera 32 No 22-08 Santa Marta, Colombia}
\newcommand{\MSU}{Department of Physics and Astronomy, Michigan State 
University, East Lansing, Michigan 48824, USA}
\newcommand{\Crookston}{Math, Science and Technology Department, University 
of Minnesota Crookston, Crookston, Minnesota 56716, USA}
\newcommand{\Duluth}{Department of Physics and Astronomy, 
University of Minnesota Duluth, Duluth, Minnesota 55812, USA}
\newcommand{\Minnesota}{School of Physics and Astronomy, University of 
Minnesota Twin Cities, Minneapolis, Minnesota 55455, USA}
\newcommand{\Mississippi}{University of Mississippi, University, Mississippi 38677, USA}
\newcommand{\NISER}{National Institute of Science Education and Research, An OCC of Homi Bhabha
National Institute, Bhubaneswar, Odisha, India}
\newcommand{\OSU}{Department of Physics, Ohio State University, Columbus,
Ohio 43210, USA}
\newcommand{\Oxford}{Subdepartment of Particle Physics, 
University of Oxford, Oxford OX1 3RH, United Kingdom}
\newcommand{\Panjab}{Department of Physics, Panjab University, 
Chandigarh, 160 014, India}
\newcommand{\Pitt}{Department of Physics, 
University of Pittsburgh, Pittsburgh, Pennsylvania 15260, USA}
\newcommand{\QMU}{Particle Physics Research Centre, 
Department of Physics and Astronomy,
Queen Mary University of London,
London E1 4NS, United Kingdom}
\newcommand{\RAL}{Rutherford Appleton Laboratory, Science 
and 
Technology Facilities Council, Didcot, OX11 0QX, United Kingdom}
\newcommand{\SAlabama}{Department of Physics, University of 
South Alabama, Mobile, Alabama 36688, USA} 
\newcommand{\Carolina}{Department of Physics and Astronomy, University of 
South Carolina, Columbia, South Carolina 29208, USA}
\newcommand{\SDakota}{South Dakota School of Mines and Technology, Rapid 
City, South Dakota 57701, USA}
\newcommand{\SMU}{Department of Physics, Southern Methodist University, 
Dallas, Texas 75275, USA}
\newcommand{\Stanford}{Department of Physics, Stanford University, 
Stanford, California 94305, USA}
\newcommand{\Sussex}{Department of Physics and Astronomy, University of 
Sussex, Falmer, Brighton BN1 9QH, United Kingdom}
\newcommand{\Syracuse}{Department of Physics, Syracuse University,
Syracuse NY 13210, USA}
\newcommand{\Tennessee}{Department of Physics and Astronomy, 
University of Tennessee, Knoxville, Tennessee 37996, USA}
\newcommand{\Texas}{Department of Physics, University of Texas at Austin, 
Austin, Texas 78712, USA}
\newcommand{\Tufts}{Department of Physics and Astronomy, Tufts University, Medford, 
Massachusetts 02155, USA}
\newcommand{\UCL}{Physics and Astronomy Department, University College 
London, 
Gower Street, London WC1E 6BT, United Kingdom}
\newcommand{\Virginia}{Department of Physics, University of Virginia, 
Charlottesville, Virginia 22904, USA}
\newcommand{\WSU}{Department of Mathematics, Statistics, and Physics,
 Wichita State University, 
Wichita, Kansas 67260, USA}
\newcommand{\WandM}{Department of Physics, William \& Mary, 
Williamsburg, Virginia 23187, USA}
\newcommand{\Wisconsin}{Department of Physics, University of 
Wisconsin-Madison, Madison, Wisconsin 53706, USA}
\newcommand{\deceased}{Deceased.}
\affiliation{\ANL}
\affiliation{\Atlantico}
\affiliation{\Bandirma}
\affiliation{\BHU}
\affiliation{\Caltech}
\affiliation{\Charles}
\affiliation{\Cincinnati}
\affiliation{\Cochin}
\affiliation{\CSU}
\affiliation{\CTU}
\affiliation{\Delhi}
\affiliation{\Erciyes}
\affiliation{\FNAL}
\affiliation{\FSU}
\affiliation{\UFG}
\affiliation{\Guwahati}
\affiliation{\Houston}
\affiliation{\Hyderabad}
\affiliation{\IHyderabad}
\affiliation{\IIT}
\affiliation{\Imperial}
\affiliation{\Indiana}
\affiliation{\ICS}
\affiliation{\INR}
\affiliation{\IOP}
\affiliation{\UIowa}
\affiliation{\ISU}
\affiliation{\Irvine}
\affiliation{\JINR}
\affiliation{\Magdalena}
\affiliation{\MSU}
\affiliation{\Duluth}
\affiliation{\Minnesota}
\affiliation{\Mississippi}
\affiliation{\NISER}
\affiliation{\OSU}
\affiliation{\Panjab}
\affiliation{\Pitt}
\affiliation{\QMU}
\affiliation{\SAlabama}
\affiliation{\Carolina}
\affiliation{\SMU}
\affiliation{\Sussex}
\affiliation{\Syracuse}
\affiliation{\Texas}
\affiliation{\Tufts}
\affiliation{\UCL}
\affiliation{\Virginia}
\affiliation{\WSU}
\affiliation{\WandM}
\affiliation{\Wisconsin}

\author{S.~Abubakar}
\affiliation{\Erciyes}

\author{M.~A.~Acero}
\affiliation{\Atlantico}

\author{B.~Acharya}
\affiliation{\Mississippi}

\author{P.~Adamson}
\affiliation{\FNAL}









\author{N.~Anfimov}
\affiliation{\JINR}


\author{A.~Antoshkin}
\affiliation{\JINR}


\author{E.~Arrieta-Diaz}
\affiliation{\Magdalena}

\author{L.~Asquith}
\affiliation{\Sussex}


\author{A.~Aurisano}
\affiliation{\Cincinnati}

\author{N.~Balashov}
\affiliation{\JINR}

% ...

\collaboration{The NOvA Collaboration}
\noaffiliation

\date{\today}
\begin{abstract}
NOvA has reported a significance of mass ordering determination using ten years of data together with external constraints from reactor-based experiments.  The JUNO collaboration is poised to provide a more precise reactor-based constraint on $|\Delta m^2_{32}|$.  In this Letter, we explore the potential impact of this anticipated measurement on the determination of the neutrino mass ordering by NOvA.  We find that $3\sigma$ evidence of the normal ordering is achievable over a range of plausible JUNO measurements within the next five years.
\end{abstract}

\maketitle
Neutrinos are subject to quantum mechanical mixing of their flavor states as they propagate~\cite{Pontecorvo:1957cp,Pontekorvo1978}.  The probability that a neutrino created with one flavor is later detected with a different flavor depends on the neutrino energy $E$ and the distance it travels $L$, as well as on the differences in the squared masses of the neutrino mass states $\Delta m^2_{ij}=m^2_i-m^2_j$; three mixing angles labeled $\theta_{12},$ $\theta_{23},$ and $ \theta_{13}$; and a phase $\delta_{\rm CP}$.  If $\delta_{\rm CP}\ne0~{\rm or}~\pi$, neutrinos violate charge-parity (CP) symmetry. The mixing angles and $\delta_{\rm CP}$ parameterize the unitary PMNS matrix, which relates the neutrino mass eigenstates to the flavor eigenstates~\cite{Maki:1962mu}. Importantly, the oscillation probability additionally depends on the ordering of the neutrino masses, a significant effect when neutrinos propagate long distances in matter~\cite{msw_2005}.  While it is known that $\Delta m^2_{21}\sim+8\times10^{-5}{\rm~eV^2}$ and that $|\Delta m^2_{32}|\sim2\times10^{-3}{\rm~eV^2}$, the sign of the latter remains unknown~\cite{ParticleDataGroup:2022pth}.
Positive $\Delta m^{2}_{32}$ is called the normal ordering, while negative $\Delta m^2_{32}$ is called the inverted ordering. 

Determining the mass ordering is crucial to understanding the nature of neutrino mass~\cite{King:2014nza}, interpreting supernova neutrino fluxes~\cite{Scholberg:2017czd}, deciphering results from neutrinoless double beta decay experiments~\cite{Dolinski:2019nrj}, and informing cosmological fits~\cite{Lesgourgues_2012}.  There are multiple observations that have bearing on the neutrino mass ordering, yet the question is unsettled.  Previous results from NOvA~\cite{NOvA:2025tmb}, T2K~\cite{T2K:2025yoy}, and a joint Super-Kamiokande and T2K analysis~\cite{T2K:2024wfn} exhibit mild preferences for normal ordering.  Cosmological data place limits on the sum of neutrino masses that disfavor the inverted mass ordering~\cite{DESI:2024hhd,SPT-3G:2024atg}.  On the other hand, the joint NOvA-T2K oscillation analysis finds that good agreement for the CP-violating phase is only achieved with the inverted mass ordering hypothesis~\cite{T2K:2025wet}.  In this Letter, we report the significance of the resolution of the mass ordering using 10 years of NOvA data, constrained by plausible future measurements of $\Delta m^2_{32}$ from a reactor experiment with the sensitivity of JUNO~\cite{JUNOSensitivity}.

\begin{figure*}[t]
 \includegraphics[width=.85\textwidth]{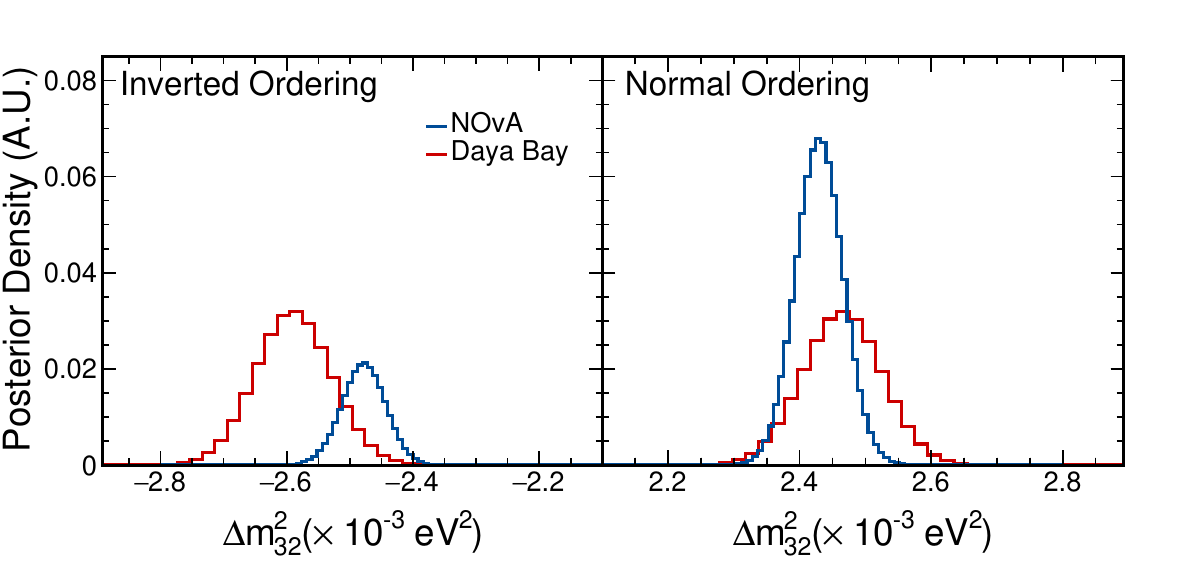}
 \caption{Posterior probability density distributions (in arbitrary units) of $\Delta m^2_{32}$ measured by NOvA~\cite{NOvA:2025tmb} and Daya Bay~\cite{DayaBay:2022orm} under the inverted (left) and normal (right) mass-ordering hypotheses. Greater consistency of the distributions under the normal ordering hypothesis strengthens the preference for the normal ordering beyond the significance derived from a NOvA-only fit.}
 \label{fig:novadbdm32}
\end{figure*}

As detailed in Refs.~\cite{PhysRevD.72.013009,Parke:2024xre}, the wavelength of neutrino oscillation in $L/E$ 
is given by a weighted combination of the mass-squared differences. This weighted combination depends on the sign of the $\Delta m^2$ values, not just the magnitude.  Moreover, the weights in the combination depend on the neutrino flavor that is disappearing.  Experiments based at nuclear reactors observe $\overline{\nu}_{e}$ disappearance, while experiments that study accelerator neutrino oscillations over a long baseline observe $\nu_{\mu}$ and $\overline{\nu}_{\mu}$ disappearance.
If the correct ordering is assumed, the values of $\Delta m^2_{32}$ (or equivalently $\Delta m^2_{31}$) extracted from the reactor and long-baseline measurements should agree, while if the incorrect ordering is assumed the values would disagree.  With only a few percent difference expected in the inferred values of the mass squared differences in the wrong ordering, reactor and long-baseline experiments must measure $\Delta m^2_{32}$  (or $\Delta m^2_{31}$) to $\sim 1\%$ to be sensitive to the difference.  

Long-baseline experiments also observe $\nu_{\mu}\rightarrow\nu_{e}$ appearance for both neutrinos and antineutrinos.  In matter,  the coherent interaction between the electron neutrino component of the propagating neutrinos and the electrons in the earth introduces a mass ordering dependence in the appearance probability~\cite{MIKHEYEV198941,Wolfenstein:1977ue}.  In one ordering, the electron neutrino appearance probability is enhanced and the electron antineutrino appearance probability is suppressed relative to vacuum, while the opposite happens in the other mass ordering.  This asymmetry is the basis of ordering measurements in long-baseline experiments, but CP violation has a similar effect, enhancing neutrino appearance probability and suppressing antineutrino appearance probability, or vice versa, depending on the value of $\delta_{\rm CP}$.    

NOvA is an accelerator neutrino oscillation experiment with an \unit[810]{km} baseline and an average neutrino energy of \unit[2]{GeV}.  With this long baseline, NOvA is sensitive to the mass ordering, but this sensitivity depends strongly on $\delta_{\rm CP}$.  In a recently published Letter~\cite{NOvA:2025tmb}, we reported results obtained using 10 years of $\nu_{\mu}$ disappearance and $\nu_{e}$ appearance data for both neutrinos and antineutrinos. In a Bayesian inference framework, NOvA data without external constraints on $\theta_{13}$ or $\Delta m^2_{32}$ yield a posterior probability of 70\% favoring the normal ordering over the inverted ordering. When external constraints on $\Delta m^2_{32}$ and $\sin^2\theta_{13}$ from the Daya Bay experiment~\cite{DayaBay:2022orm} are included, the posterior probability for the normal ordering rises to 87\%, corresponding to a 1.6$\sigma$ preference. This enhanced preference arises from the complementarity between long-baseline and reactor measurements of $|\Delta m^{2}_{32}|$.
NOvA measures $\Delta m^2_{32} = 2.431^{+0.036}_{-0.034},(-2.479^{+0.036}_{-0.036}) \times 10^{-3}~\mathrm{eV}^2$ in the normal (inverted) ordering, while Daya Bay measures $\Delta m^2_{32} = 2.466^{+0.060}_{-0.060},(-2.571^{+0.060}_{-0.060}) \times 10^{-3}~\mathrm{eV}^2$~\cite{DayaBay:2022orm}.  
As shown in Fig.~\ref{fig:novadbdm32}, the two measurements are more consistent under the normal ordering hypothesis, leading to a stronger combined posterior preference for that ordering.

\begin{figure*}
\includegraphics[width=.8\linewidth]{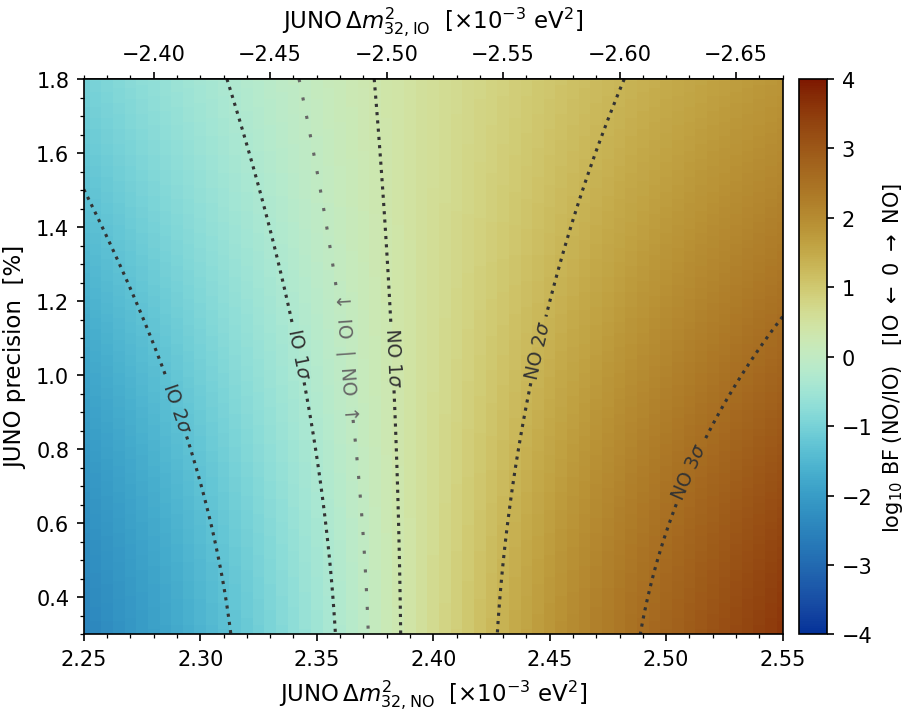}
 \caption{Sensitivity of NOvA to the neutrino mass ordering with projected JUNO constraints on $\Delta m^2_{32}$. The bottom (top) horizontal axis scans values of $\Delta m^2_{32}$ assuming normal (inverted) ordering, with a fixed offset of $0.12 \times 10^{-3}~\mathrm{eV}^2$ between the two. The vertical axis gives the fractional precision on $\Delta m^2_{32}$. The color scale indicates the Bayes factor (BF) on a log scale, where positive (negative) values favor normal (inverted) ordering. Iso-contours trace BF values corresponding to the $1,~2,~{\rm and }~3\sigma$ credible intervals.}
 \label{fig:mosensitivity}
\end{figure*}

With a precision of 2.4\%, Daya Bay currently provides the best determination of $|\Delta m^2_{32}|$ from electron antineutrino disappearance; however, JUNO is expected to achieve sub-percent precision on this mass squared splitting~\cite{JUNOSensitivity}. JUNO began data taking in August 2025 and reported first results in November 2025~\cite{newjuno}, already reaching world-leading precision on the solar (12-sector) oscillation parameters. A determination of $|\Delta m^{2}_{32}|$ is anticipated in a forthcoming publication. With six to seven years of exposure, JUNO is projected to resolve the mass ordering at the $3\sigma$ level on its own~\cite{JUNOSensitivity}.

Here, we assess the impact that a more precise constraint on $\Delta m^2_{32}$ from an electron antineutrino disappearance measurement would have on the near term determination of the neutrino mass ordering. For this analysis the NOvA event selection criteria, systematic treatment, and other methodological details follow those reported in Ref.~\cite{NOvA:2025tmb}.
The analysis is updated to include solar parameter constraints from JUNO~\cite{newjuno}, with negligible impact on the posterior distributions relative to Ref.~\cite{NOvA:2025tmb}.
Oscillation probabilities are computed with \textsc{NuFast}~\cite{PhysRevD.110.073005}, and posteriors are sampled using NOvA's implementation of Hamiltonian Markov chain Monte Carlo~\cite{NOvA:2023iam} using Stan~\cite{stan2026}.
We retain the external constraint on $\sin^2(2\theta_{13}) = 0.0851 \pm 0.0024$ from Daya Bay~\cite{DayaBay:2022orm}, consistent with the treatment in the JUNO analysis~\cite{newjuno}.
As shown in Ref.~\cite{JUNOSensitivity}, the correlation between $\sin^2(2\theta_{13})$ and $\Delta m^2_{32}$ in JUNO is negligible, allowing these parameters to be treated as independent in this analysis.

The $\Delta m^2_{32}$ constraint from Daya Bay is removed and replaced with a series of hypothetical reactor constraints. We then recompute the mass ordering significance across a range of representative central values of $\Delta m^2_{32}$ and associated fractional uncertainties up to 2\%.  The hypothetical constraint is modeled as a one-dimensional bi-Gaussian prior in $\Delta m^2_{32}$, with peaks centered at values corresponding to the normal and inverted ordering hypotheses. The two densities are assigned equal prior weight, and their widths reflect the assumed experimental precision.
The pairs of values of $|\Delta m^2_{32}|$ under the two orderings are separated by $\lambda\equiv\delta(|\Delta m^2_{32}|)_{\mathrm{NO-IO}}$, an offset that depends on the solar oscillation parameters $\Delta m^{2}_{21}$ and $\sin^2\theta_{12}$ and is a function of $E$ for a given baseline $L$~\cite{parke2}.  The offset observed by JUNO will additionally depend on the detector energy resolution, energy response nonlinearity, and statistical fluctuations, particularly at low exposure.  At JUNO’s flux-averaged energy, baseline, and anticipated detector performance, this offset is estimated to be $\lambda \sim 0.12 \times 10^{-3}~\mathrm{eV}^2$~\cite{Parke:2024xre, cabrera}. In this work, we adopt that estimate and provide a procedure to adjust results to other reasonable values of this offset
($\sim \pm 0.005 \times 10^{-3}~\mathrm{eV}^2$ relative to our chosen value).

NOvA's four-dimensional posterior in $\delta_{\rm CP},$ $\sin^2\theta_{23},$ $\sin^2(2\theta_{13}),$ and $\Delta m^2_{32}$ is marginalized over all systematic parameters and the hypothetical reactor constraint to extract the Bayes factor, which quantifies the relative preference for one mass ordering over another. The number of sigma significance corresponding to a given Bayes factor is found by mapping the posterior probability to the two-sided Gaussian tail probability. For example, with equal prior odds, a 90\% posterior probability corresponds to a Bayes factor of 9, whereas a 3$\sigma$ two-sided confidence interval corresponds to a Bayes factor of roughly 370.

\begin{figure}[htb]
  \centering
  \includegraphics[width=\linewidth]{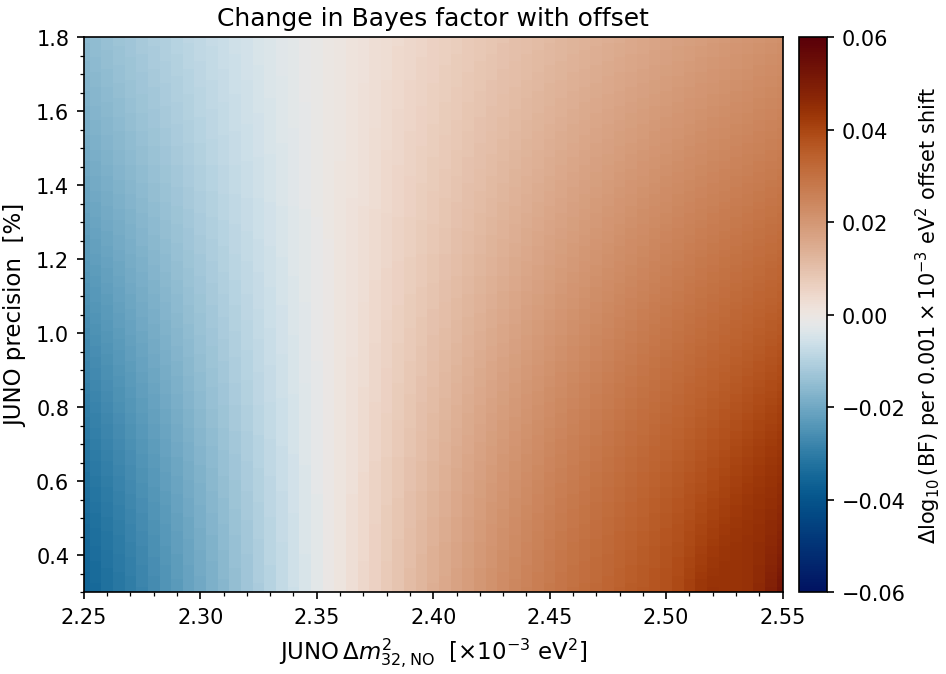}
  \caption{The function $\frac{d[\log_{10}(\rm{BF})]}{d\lambda}$ evaluated at $\lambda_0=0.12\mathord{\times}10^{-3}\ \rm{eV}^2$, allowing adjustment to the Fig. 2 results based on the actual offset seen in JUNO's measurements.  The horizontal axis scans $\Delta m^2_{32}$ under normal ordering, and the vertical axis shows the fractional precision. The color scale gives the change in $\log_{10}\rm{(BF)}$ per $0.001\times10^{-3} {\rm eV^{2}}$ change in $\lambda$ relative to the nominal value used in Fig.~\ref{fig:mosensitivity}.}
 \label{fig:mosensitivity_slope}
\end{figure}

Figure~\ref{fig:mosensitivity} shows the logarithm of the Bayes factor (BF) for the neutrino mass ordering preference extracted from the NOvA analysis as a function of the hypothetical $\Delta m^2_{32}$ and its fractional precision as measured by the reactor experiment. The parameters $\Delta m^2_{32,\mathrm{NO}}$ (bottom horizontal axis) and $\Delta m^2_{32,\mathrm{IO}}$ (top horizontal axis) are related by the offset of $0.12 \times 10^{-3}~\mathrm{eV}^2$. The Bayes factor is defined as the ratio of posterior probabilities for normal versus inverted ordering; thus, $\log_{10}(\mathrm{BF})>0$ ($<0$) indicates a preference for normal (inverted) ordering, while $\log_{10}(\mathrm{BF})=0$ corresponds to no preference. The isocontours corresponding to the equivalent of $1\sigma$, $2\sigma$, and $3\sigma$ significances are overlaid on the plot. 

As discussed, the offset $\lambda$ measured by JUNO need not coincide with the value used to compute the $\log_{10}(\mathrm{BF})$ values in Fig.~\ref{fig:mosensitivity}. In Fig.~\ref{fig:mosensitivity_slope}, we show the change in $\log_{10}(\mathrm{BF})$ as a function of the deviation of JUNO's $\lambda$ from the reference value $\lambda_0 = 0.12\times 10^{-3}{\rm~eV^2}$ used in this analysis, i.e., $\left.\frac{d[\log_{10}(\mathrm{BF})]}{d\lambda}\right|_{\lambda_0}$. The Bayes factor at a different offset can then be calculated as $\left.\log_{10}(\mathrm{BF}(\lambda)) \simeq \log_{10}(\mathrm{BF}(\lambda_0)) + \frac{d[\log_{10}(\rm{BF})]}{d\lambda}\right|_{\lambda_0} \times (\lambda - \lambda_0)$. 
This approximation works well across the range of lambda values expected in JUNO measurements,
 as these shifts from $\lambda_0$ are small relative to NOvA's experimental resolution on $|\Delta m^2_{32}|$.
The data release containing the values shown in Figs.~\ref{fig:mosensitivity} and ~\ref{fig:mosensitivity_slope} is provided in the supplementary material.

NOvA data analyzed in combination with reactor measurements yielding $\Delta m^2_{32} > 2.490 \times 10^{-3}~\mathrm{eV}^2$ in the normal ordering and the corresponding $\Delta m^2_{32} < -2.610 \times 10^{-3}~\mathrm{eV}^2$ in the inverted ordering with a precision of 0.3\% or better would provide evidence for the normal mass ordering at $>3\sigma$ significance. As the assumed central value increases, the same significance can be achieved with modestly reduced precision, as indicated by the isocontours in Fig.~\ref{fig:mosensitivity}.
However, when $\Delta m^2_{32} > 2.55 \times 10^{-3}~\mathrm{eV}^2$, measurements from NOvA and the reactor experiment would be in strong tension under both orderings, rendering the combined inference of mass ordering internally inconsistent and therefore unreliable.  To find significant evidence for the inverted ordering, the reactor measurement would have to be less than $2.25\times10^{-3}~{\rm eV^2}$ in the normal ordering, a value that is much lower than the range of values for that parameter measured by any experiment. 

For a reactor measurement with better than $\sim 0.3$\% precision, the significance of the mass ordering determination is limited by the precision of the NOvA data.  Even small improvements in the NOvA uncertainty on the mass squared splitting, beyond the current 1.5\%, increase the range of potential reactor measurements for which a $3\sigma$ mass ordering resolution is possible.  Further improvement to the NOvA measurement requires additional muon neutrino disappearance data and tighter control of systematic uncertainties, both of which are expected before the end of NOvA.

In conclusion, we have demonstrated the impact of a precise reactor measurement of $\Delta m^2_{32}$ on NOvA's determination of the mass ordering.  A future precision reactor measurement could enhance NOvA's current $1.6\sigma$ preference and potentially provide evidence for the normal mass ordering of neutrinos.  If, however, the reactor central value of $\Delta m^2_{32}$ remains near the current global average, the precision anticipated from JUNO alone would be insufficient to enable an early determination of mass ordering through the complementarity of reactor and long-baseline $\Delta m^2_{32}$ measurements. In that scenario, a significant result would require either improved precision from long-baseline measurements or the accumulation of sufficient exposure for one or more experiments to establish the ordering independently.

\section*{Acknowledgements}

\begin{acknowledgments}
This document was prepared by the NOvA collaboration using the resources of the Fermi National Accelerator Laboratory (Fermilab), a U.S. Department of Energy, Office of Science, HEP User Facility. Fermilab is managed by Fermi Forward Discovery Group, LLC, acting under Contract No. 89243024CSC000002.  
This research used resources of the National Energy Research Scientific Computing Center, a DOE Office of Science User Facility supported by the Office of Science of the U.S. Department of Energy under Contract No. DE-AC02-05CH11231 using NERSC award HEP-ERCAP0028967.
This work was supported by the U.S. Department of Energy; the U.S. National Science Foundation; the Department of Science and Technology, India; the European Research Council; the MSMT CR, GA UK, Czech Republic; the RAS, the Ministry of Science and Higher Education, and RFBR, Russia; CNPq and FAPEG, Brazil; UKRI, STFC and the Royal Society, United Kingdom; and the State and University of Minnesota.  We are grateful for the contributions of the staffs of the University of Minnesota at the Ash River Laboratory, and of Fermilab. For the purpose of open access, the authors have applied a Creative Commons Attribution (CC BY) license to any Author Accepted Manuscript version arising.

\end{acknowledgments}

\FloatBarrier
\bibliographystyle{apsrev4-1}
\bibliography{cites}

\clearpage

\end{document}